\documentclass[reprint, amsmath, amssymb, aps, prl]{revtex4-1}
\usepackage{graphicx}
\usepackage{dcolumn}
\usepackage{tensor}
\usepackage{amssymb}
\usepackage{bm}
\usepackage{amsmath,mleftright}
\usepackage{xparse}

\NewDocumentCommand{\evalat}{sO{\big}mm}{%
  \IfBooleanTF{#1}
   {\mleft. #3 \mright|_{#4}}
   {#3#2|_{#4}}%
}

\begin{document}
\preprint{APS/123-QED}
\title{Optical magnetic dipole levitation using a plasmonic surface}
\author{Jack J. \surname{Kingsley-Smith}}
\author{Michela F. \surname{Picardi}}
\author{Francisco J. \surname{Rodr\'iguez-Fortu\~no}}
\thanks{francisco.rodriguez\_fortuno@kcl.ac.uk}
\affiliation{Department of Physics, King's College London, Strand, London WC2R 2LS, United Kingdom}
\date{\today}

\begin{abstract}
\noindent Optically-induced magnetic resonances in non-magnetic media have unlocked magnetic light-matter interactions and led to new technologies in many research fields. 
Previous proposals for the levitation of nanoscale particles without structured illumination have worked on the basis of epsilon-near-zero surfaces or anisotropic materials but these carry with them significant fabrication difficulties. 
We report the optical levitation of a magnetic dipole over a wide range of realistic materials, including bulk metals, thereby relieving these difficulties. The repulsion is independent of surface losses and we propose an experiment to detect this force which consists of a core-shell nanoparticle, exhibiting a magnetic resonance, in close proximity to a gold substrate under plane wave illumination. We anticipate the use of this phenomenon in new nanomechanical devices. 
\end{abstract}

\maketitle

\noindent Levitation is one of the most exciting phenomena in modern physics and finds many applications in cutting-edge technologies. From locomotive maglev systems to superconducting magnets \cite{Brandt1989,Ma2003,Lee2006}, repulsive forces have been a part of many breakthroughs in modern engineering. 
At the microscopic scale, optical tweezers \cite{Ashkin1986,Ashkin2000} have been the go-to tool for levitating systems for decades because of their ability to noninvasively trap a wide variety of bodies in a 3D potential well. Optomechanical systems manipulate this potential well to control the motion of the targets and provide a net force in a desired direction \cite{Grier2003,Neuman2004}. However, the continual miniaturization of such systems means that levitated objects are inevitably coming into close proximity with surfaces and the objects themselves are shrinking to the subwavelength regime. In these scenarios, conventional optical tweezers can suffer from complex surface effects \cite{Schaffer2007} and struggle to reach the trapping potentials needed to manipulate very small particles \cite{Juan2011}. 

Subwavelength particles are at the forefront of modern nanotechnology and are now widely used in physics, chemistry, biology and medicine \cite{Zeng2011,Ruedas-Rama2012,Jahani2016,Bobo2016,Amendola2017} because of their unique optical resonances \cite{Hutter2004,Maier2007} and ease of functionalization \cite{Mout2012}. 
Prior nanomechanial studies have shown that polarized particles acting as an electric dipole may be repelled from engineered materials such as epsilon-near-zero (ENZ) \cite{Rodriguez-Fortuno2014,Krasikov2014,Rodriguez-Fortuno2016}, two-dimensional \cite{Rodriguez-Fortuno2018} and multilayer \cite{Wang2016} materials. Such engineering significantly complicates the real-world applicability of these techniques. 

The work presented here differs from these approaches by utilizing optically induced magnetic dipole resonances \cite{Evlyukhin2012}, thus revealing new opportunities for levitation. Optical magnetic resonances do not necessarily require magnetic materials \cite{Pendry1999,Smith2000,Shelby2001,Shalaev2007} and have been recently studied, in conjunction with electric dipoles, with high refractive index nanoparticles \cite{Krasnok2012,Evlyukhin2012,Fu2013,Zywietz2014,Permyakov2015,Wozniak2015,Kuznetsov2016,Picardi2019a} or similar sources \cite{Evlyukhin2011,Rolly2012,Hancu2014,Alaee2015}, in the context of directional scattering \cite{Picardi2017a,Abdelrahman2019}, optical tractor beams \cite{Chen2011,Petrov2016} and Huygens metasurfaces \cite{Pfeiffer2013,Decker2015}, among others. The resonances can also be enhanced and tuned by precisely designing the scattering body \cite{Feng2017,Wei2016}. The rise of induced magnetic dipoles have enabled experimental magnetic light-matter interactions and have proven to be an excellent tool in building novel photonic systems. 

In this letter, we propose the levitation of any magnetic dipole source over a simple isotropic surface with realistic permittivities. 
The constraints on the force are shown to be surprisingly robust and realistic with a broadband repulsion expected over most metals and an independence with respect to substrate material losses, which are often the crux of near-field plasmonic technologies. We numerically demonstrate this force with a core-shell nanoparticle near a bulk gold surface, which represents an optically-induced magnetic dipole source when under plane wave illumination. This system is realistic given modern fabrication techniques \cite{Wang2018} and is merely a single example of potential illumination-source-surface combinations that would produce this effect, leading us to expect that this work will inspire new levitating technologies in a wide range of fields. 

\begin{figure}
    \centering
    \includegraphics[width=0.4\textwidth]{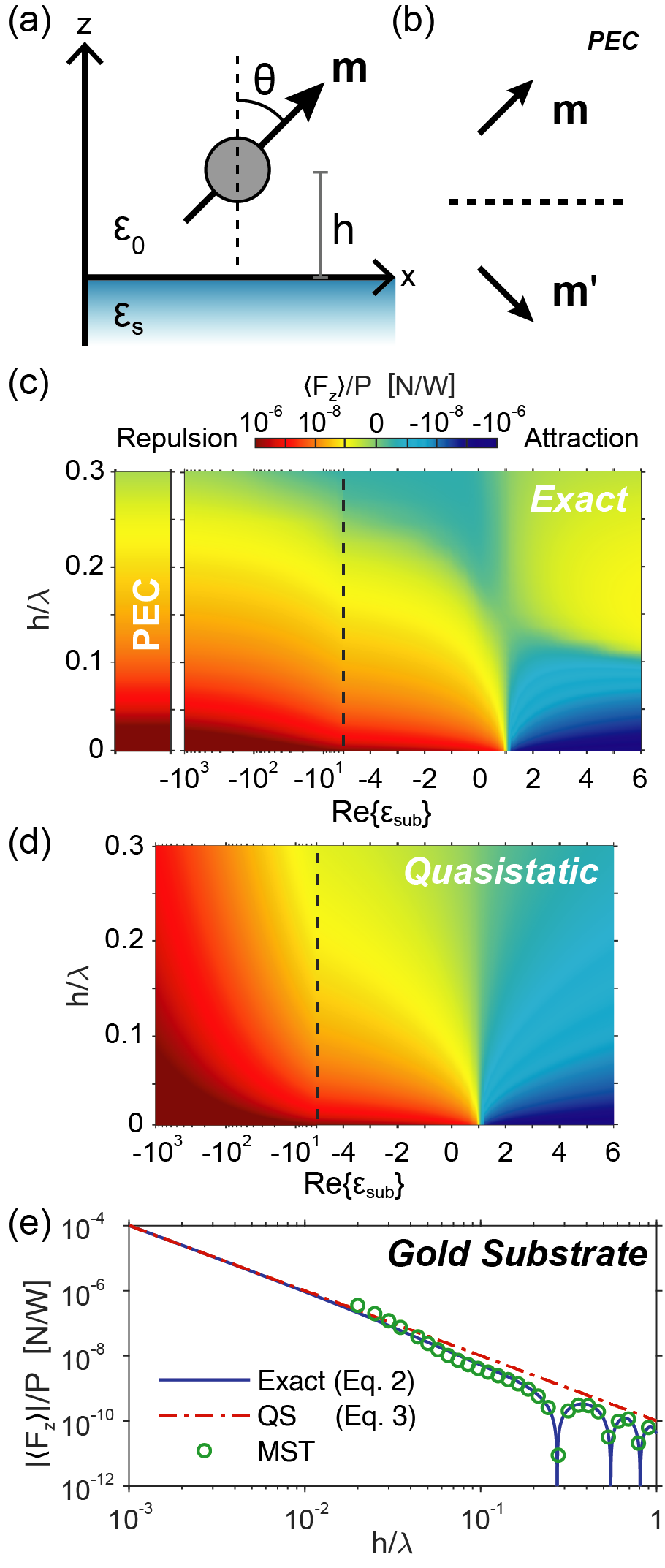}
    \caption{(a) A magnetic dipole $\mathbf{m}$ at a height $h$ above a flat isotropic surface of permittivity $\varepsilon_{\text{s}}$. The $\theta = 0$ case corresponds to the VMD. (b) For a perfect electric conducting substrate, the dipole is repelled an image dipole where the vertical component is flipped. (c) The optical force on a VMD above a near-lossless surface of $\varepsilon_{\text{s}} = \mathfrak{Re}\{\varepsilon_{\text{s}}\} + 0.01i$, calculated from Eq.\,(\ref{eq:mzforce}). The dipole is repelled from the surface for $\mathfrak{Re}\{\varepsilon_{\text{s}}\} <1$ when $h<0.2\lambda$. The left strip indicates the force over a perfect electric conductor. Red and blue indicate repulsion and attraction towards the surface, respectively. (d) The quasistatic limit of the force, given by Eq.\,(\ref{eq:QSforce}), used to model $\langle F_z\rangle$ near a realistic metal. (e) The absolute vertical force on a VMD, normalized by the radiative power, radiating at 633 nm over a gold \cite{Palik1998} surface with respect to height (normalized by the dipole wavelength). `QS' and `MST' denote the quasistatic and Maxwell stress tensor \cite{Novotny2006} calculations of the force, respectively.}
    \label{fig:mzdiagram}
\end{figure}

We begin by considering a remarkably simple scenario with a time-harmonic vertical magnetic dipole (VMD) $\mathbf{m} = (0,0,m_z)$, oscillating at an angular frequency $\omega$, in vacuum at a height $h$ above a semi-infinite isotropic substrate characterized by a relative complex permittivity $\varepsilon_{\text{s}}$, as depicted in Fig.\,\ref{fig:mzdiagram}a. This scenario fixes the tilt angle of the dipole to $\theta = 0^\circ$ for simplicity in the analytical expressions but the repulsive behavior remains in most cases for large values of $\theta$, which we show in the supplementary material \cite{SI}.

The time-averaged optical force on a time-harmonic magnetic dipole in free space near a surface is given by \cite{Nieto-Vesperinas2010,Kingsley-Smith2018}
\begin{equation}\label{equ:DipolarForce}
\langle \mathbf{F} \rangle = \frac{\mu_0}{2} \mathfrak{Re}\big\{ (\mathbf{\nabla} \otimes \mathbf{H})\, \mathbf{m}^*\big\},
\end{equation}
where $\mu_0$ denotes the free space permeability, $\mathbf{H}$ denotes the magnetic field acting on the dipole (corresponding to the reflection from the surface), an asterisk represents complex conjugation and $\otimes$ denotes an outer product. 

The reflected fields of a dipole can be expressed in the angular representation \cite{Novotny2006,Picardi2017,Kingsley-Smith2018} as an integral over the transverse wavevector $k_t$. By substituting this known expression into Eq.\,(\ref{equ:DipolarForce}) and analytically integrating over the angular dependence, we arrive at the expression for the optical force on a VMD near a planar surface (the full derivation is given in the supplementary \cite{SI}) 
\begin{equation}\label{eq:mzforce}
    \evalat{\langle F_z \rangle}{\theta = 0^\circ} = - |m_z|^2 \frac{\mu_0 }{8 \pi} \mathfrak{Re}  \bigg\{ \int_0^\infty e^{2i h k_{z}}  k_t^3 \,  r_s \, dk_t\bigg\},
\end{equation}
\noindent where $k_z = \sqrt{k_0^2-k_t^2}$ and $k_0$ is the wavenumber in the upper-half space. All properties of the surface are contained within the Fresnel reflection coefficients which we denote as $r_s$ and $r_p$ for $s$ and $p$ polarization, respectively. It is important to note that a VMD emits purely $s$-polarized light so only the $r_s$ coefficient appears in Eq.\,(\ref{eq:mzforce}). This is ultimately the key to why a magnetic dipole levitates and an electric dipole does not, as will become clear soon. No electric dipole emits purely $s$-polarized light, regardless of its polarization, as part of its radiation will always be $p$-polarized.
\begin{table}[b]
\caption{Fresnel reflection coefficients for a single interface, their quasistatic expansions and their limits when the surface is a perfect electric conductor. $k_{z1}$ and $k_{z2}$ are the $z$-component of the wavevector in the upper and lower half-space, respectively.}
\label{tab:1}
\begin{tabular}{|c|c|c|c|}
\hline
   & General Form & Quasistatic Form ($k_t\gg k_0)$ & PEC \\ \hline
$r_s$ & $\frac{k_{z1} -  k_{z2}}{k_{z1} +   k_{z2}}$  & $0 + k_0^2\,\frac{\varepsilon_{\text{s}} - 1}{4 \, k_{t}^2} + \mathcal{O}\bigg(\frac{1}{k_{t}^4}\bigg)$       & $-1$  \\ \hline
$r_p$ & $\frac{\varepsilon_{\text{s}} \, k_{z1} - \, k{z2}}{\varepsilon_{\text{s}} \, k_{z1} + \, k{z2}}$  & $\frac{\varepsilon_{\text{s}}-1}{\varepsilon_{\text{s}}+1} + \mathcal{O}\bigg(\frac{1}{k_{t}^2}\bigg)$      & $1$   \\ \hline
\end{tabular}
\end{table}

Throughout this work, we assume a non-magnetic substrate such that the permeability $\mu_s = 1$, so $r_s$ is a function of $\varepsilon_{\text{s}}$ and $k_t$, given by Table \ref{tab:1}, and allowing Eq.\,(\ref{eq:mzforce}) to be calculated numerically for any permittivity and height. This leads us directly to the main result of this paper: Fig.\,\ref{fig:mzdiagram}c. It shows that a dipole placed close ($h<0.2\lambda$) to the surface experiences a repulsive force when $\mathfrak{Re}\{\varepsilon_{\text{s}}\}<1$ (a plasmonic surface).  
To give more perspective, if one considers the same scenario with an electric dipole \cite{Rodriguez-Fortuno2014}, a repulsive force is only achieved when $|\varepsilon_{\text{s}}|<1$ (i.e. in the ENZ regime), which is a far more restrictive domain. In our case, where the repulsion occurs for any plasmonic material, the surface can be as simple as a bulk metal when the dipole radiates with a frequency below the metal's plasma frequency, thus providing greater real-world applicability.

To qualitatively understand this phenomenon, consider the ideal case of a dipole $\mathbf{m}$ near a perfect electric conductor (PEC). Here, we can apply image theory \cite{Balanis1997,Griffiths2007} to calculate the backscattering force on the dipole, which states that this case is equivalent to a source dipole located at $z = h$ and an image dipole $\mathbf{m}'$ located at $z = -h$ with no surface present anywhere, as depicted in Fig.\,\ref{fig:mzdiagram}b where the magnetic dipoles are schematically represented by time-harmonic arrows. A PEC is defined by $r_s = -r_p = -1$ and it is known that the image dipole in this case is orientated as $\mathbf{m}' = (m_x,m_y,-m_z)$ \cite{Balanis1997}. The repulsion is therefore understood very clearly and can be calculated exactly by substituting $r_s = -r_p = -1$ into the generalized $\theta$ version of Eq.\,(\ref{eq:mzforce}) and solving the integral analytically. These calculations are included in the supplementary material \cite{SI} and they show that, when $h\to0$, a magnetic dipole of any orientation exhibits a $h^{-4}$ height dependent repulsive force. 

This explanation is so far very similar to that of the electric dipole repulsion over ENZ materials \cite{Rodriguez-Fortuno2014}, but the magnetic case becomes slightly more complicated when moving away from a fictitious PEC material and towards realistic metals, with  finite complex permittivites and non-zero skin depth. A PEC is an excellent approximation for a metal at very low frequencies but infrared and visible frequencies require more sophisticated models such as the Drude model to remain accurate. This approach is, however, consistent with the previous explanation because the lossless Drude model of a metal at very low frequencies corresponds to $\varepsilon_s\to-\infty$ and the exact calculation of $\langle F_z\rangle$, in this limit, tends towards the PEC case, as shown in Fig.\,\ref{fig:mzdiagram}c.

The aforementioned complication arises when dealing with realistic values of plasmonic permittivities in the optical regime, where $r_s \to 0$ for non-magnetic substrates. When the dipole approaches a surface like this, the interaction can be approximated by the quasistatic limit where $k_t \gg k_0$. When considering Eq.\,(\ref{eq:mzforce}), $r_s \to 0$ might suggest that $\langle F_z \rangle =0$, but this is not the case. In fact, this regime requires looking at the next terms in the quasistatic Taylor expansion of $r_s$ (Table \ref{tab:1}), which carry with them decreasing powers of $k_t$. By substituting the dominant term of $r_s$ into Eq.\,(\ref{eq:mzforce}), and ignoring trailing terms, the integral can be evaluated analytically, giving:
\begin{equation}\label{eq:QSforce}
     \evalat{\langle F_z \rangle^{QS}}{\theta=0^\circ} = \frac{k_0^2 \,  \mu_0 \,  |m_z|^2}{128 \pi   h^2} \big(1-\mathfrak{Re}\{\varepsilon_{\text{s}}\}\big).
\end{equation}
This quasistatic expression explicitly reveals that repulsive forces will occur under the simple condition $\mathfrak{Re}\{\varepsilon_{\text{s}}\} < 1$. This condition is shown visually in Fig.\,\ref{fig:mzdiagram}d, which plots the force calculated via Eq.\,(\ref{eq:QSforce}) and shows a close agreement with the exact calculation of Eq.\,(\ref{eq:mzforce}) in Fig.\,\ref{fig:mzdiagram}c when $h/\lambda \ll 1$.

We note that Eq.\,(\ref{eq:QSforce}) has a $h^{-2}$ dependence, while the ideal PEC case showed a $h^{-4}$ dependence. Indeed, the exact force calculated from Eq.\,(\ref{eq:mzforce}) shows a transition between  $h^{-4}$ and $h^{-2}$ dependence when $h$ approaches the skin depth of the metal (see supplementary material for a detailed explanation \cite{SI}). For very low values of the permittivity $\varepsilon\to-\infty$, the skin depth tends to zero, and so only the $h^{-4}$ dependence remains, matching the PEC case. 

It is also important to stress that Eq.\,(\ref{eq:QSforce}) shows the quasistatic force is independent of substrate losses, since $\mathfrak{Im}\{\varepsilon_{\text{s}}\}$ is not present. The repulsion exists as long as the real part of $\varepsilon_s$ is smaller than 1, regardless of its imaginary part. The significant losses of real metals typically plague near-field optical techniques and so this loss independence is a highly desirable feature for practical experiments. 

Since this plasmonic repulsion is well suited to real-world optomechanical applications, we provide Fig.\,\ref{fig:mzdiagram}e which depicts the repulsive force of a 633 nm emitting VMD in air over a bulk gold surface ($\varepsilon_{\text{s}} = -11.8 + 1.2i$ \cite{Johnson1974}). The exact results from Eq.\,(\ref{eq:mzforce}) are checked with an independent force calculation method, provided by the Maxwell stress tensor (MST) method and the reflected fields of the dipole. These fields are calculated directly with the semianalytical Green's function approach that we discuss in \cite{SI} and comes from literature \cite{Novotny2006,Picardi2017,Kingsley-Smith2018}. The quasistatic limit of Eq.\,(\ref{eq:QSforce}) is plotted alongside these results to show where the approximation is valid.

\begin{figure}
    \centering
    \includegraphics[width=0.42\textwidth]{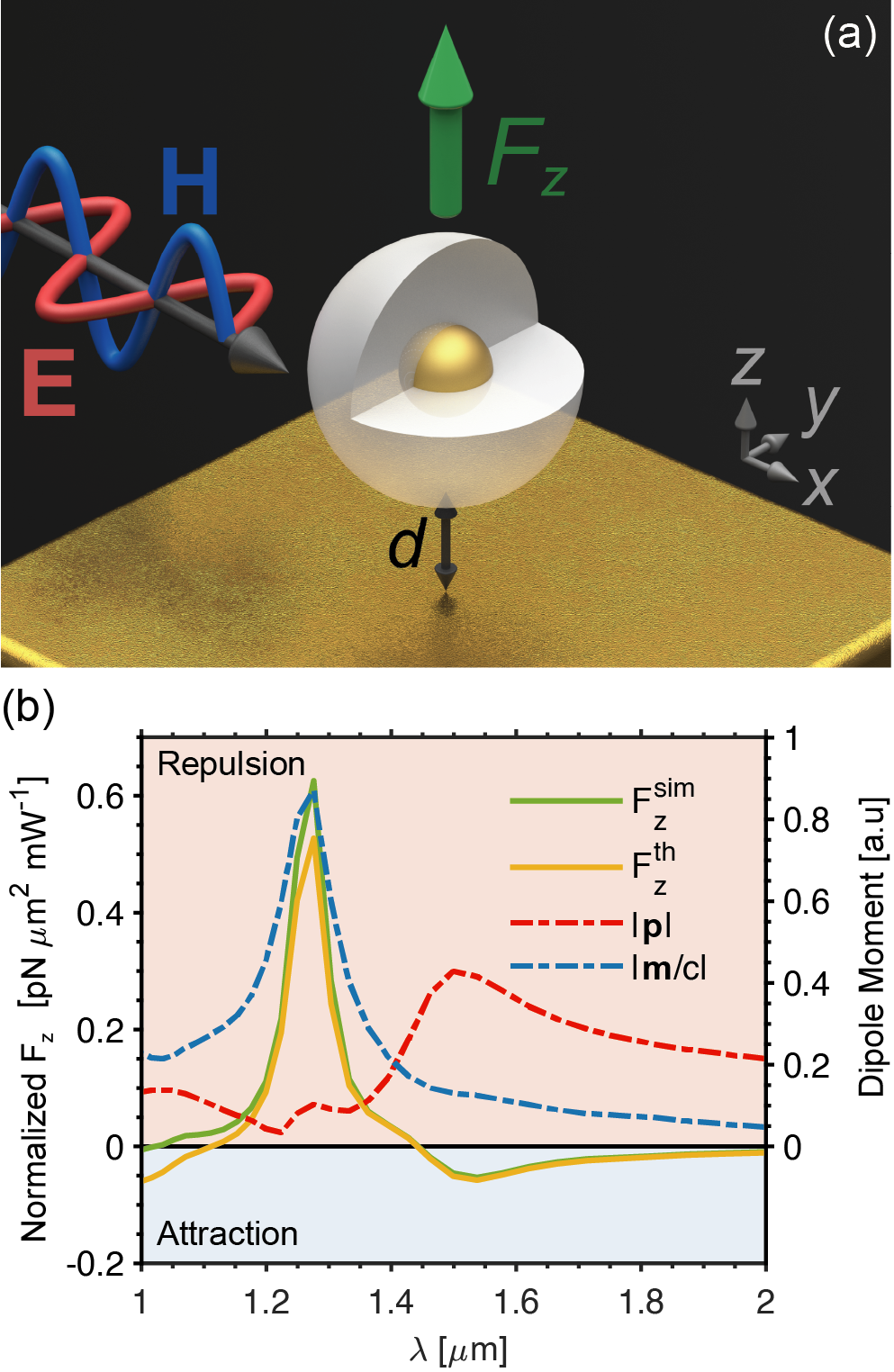}
    \caption{(a) The configuration of the force simulation, consisting of a spherical gold core-silicon shell nanoparticle with inner and outer radii of $62$ nm and $180$ nm, respectively, at a distance $d$ above a bulk gold surface. The nanoparticle is illuminated with a \textit{s}-polarized plane wave incident at $85^\circ$ to induce a magnetic dipole resonance. The particle is cross-sectioned for illustrative purposes. (b) The time-averaged vertical force, normalized by the incident plane wave power density, on the simulated nanoparticle switches between repulsive to attractive depending on the resonances of the particle as indicated by the magnitude of the induced dipole moments.}
    \label{fig:coreshell}
\end{figure}

At this point, we suggest a plausible and simple experimental realization: a core-shell nanoparticle being repelled from a gold substrate, as illustrated in Fig.\,\ref{fig:coreshell}a. The core-shell nanoparticle design from Feng \textit{et. al.} \cite{Feng2017} was adopted because it exhibits a pure magnetic dipole at 1.3 $\mu$m, meaning that the electric dipole and higher order multipoles are negligible at this frequency. This simple system is designed to demonstrate that this phenomenon can be achieved in a realistic nano-optical system. 

The particle is illuminated with a $s$-polarized plane wave incident at 85$^\circ$ which, after reflecting from the gold surface, produces a standing wave. The standing wave set up by reflection on a metal is advantageous to our purpose because the magnetic field of the standing wave is maximal at the surface, while the electric field is minimal, therefore maximizing any potential magnetic dipole resonance while reducing any potential electric dipole one. 

The magnetic repulsion has a quasistatic behavior so in order to observe an appreciable force, the particle has to be at a height $h<0.2\lambda$ above the surface \cite{Kingsley-Smith2018}. In this example, we place the particle at $d = 50$ nm so that it strongly interacts with its reflected near fields. 

In some systems, this proximity can be problematic because magnetic dipole resonances are often accompanied by a non-zero electric dipole resonance and as shown earlier, the magnetic dipole repulsion has a $h^{-2}$ dependence which contrasts to the  $h^{-4}$ dependent attraction of an electric dipole. Therefore, the electric dipole attraction can often overpower the magnetic repulsion at the required low heights. 
This is why the configuration proposed in Fig.\,\ref{fig:coreshell}a uses a particle with a magnetic response but no electric response at 1.3 $\mu$m, and why the standing wave property described above is advantageous.

We stress that this is not a unique solution to this electric attraction problem and it merely stands as a proof of the magnetic repulsion. 
The full three-dimensional electromagnetic fields of this system were calculated with a commercial electromagnetic software \cite{CST} and the simulated force from the scattered (total minus illumination) fields  $F_z^{\text{sim}}$ was calculated via Maxwell stress tensor integration, detailed in the supplementary material \cite{SI}. 

Fig.\,\ref{fig:coreshell}b confirms that the particle is repelled from the surface directly because of its magnetic dipole resonance. Away from the magnetic peak at 1.3 $\mu$m, the electric dipole attraction takes over and the net force $F_z^{\text{sim}}$ becomes negative. While the resonance peak at 1.3 $\mu$m does not illustrate the ideal magnetic dipole scattering that was desired, the residual electric dipole is still weak enough to be negligible and so the net force remains repulsive. 

As a further check, the dipole moments induced in the particle were retrieved from the far-field radiation patterns of the particle's scattering, in a procedure also described in \cite{SI}. This retrieval allows us to compare the theory with numerical simulations, by applying the general version of Eq.\,(\ref{eq:mzforce}) in \cite{SI} to produce $F_z^{\text{th}}$. The strong correlation between $F_z^{\text{sim}}$ and $F_z^{\text{th}}$ indicates that the results are self consistent and the gradual discrepancy at the high energy end of the spectrum is likely caused by emerging higher order multipole terms, which $F_z^{\text{th}}$ neglects. $F_z^{\text{sim}}$ and $F_z^{\text{th}}$ are normalized by the incident power density of the illumination. 

For comparison purposes, the forces observed with this example are comparable with the lateral forces of chiral gold particles \cite{Wang2014} that are associated with viable chiral sorting techniques. We note that the dipolar resonance spectra differs from that of the original design \cite{Feng2017} because of the polarizing effect of the backscattering radiation; the original work is absent of any reflective surfaces.


We have demonstrated a novel near-field repulsive force between an optical magnetic dipole and a planar surface without the need for magnetic or exotic materials. The repulsion can be achieved with gradientless illumination because its role is purely to excite the magnetic resonance of the levitation subject. 
The repulsive force requires a surface property of $\mathfrak{Re}\{\varepsilon_{\text{s}}\} < 1$, which corresponds to any plasmonic material, and is invariant with respect to $\mathfrak{Im}\{\varepsilon_{\text{s}}\}$, resulting in substrate loss invariance. The near-field repulsion can be explained by image theory, via an oppositely-aligned image dipole within the reflective surface, and is present for an ultra-broad range of substrate materials. A core-shell particle was then modeled over a bulk gold surface to verify this effect in a real-world setting. The repulsive force appears when the magnetic dipole resonance is dominant and closely agrees with the predicted theory. Since many common materials fit this criterion, this force neatly lends itself towards real applications, rather than just an exotic theory. This work will provide the optomechanical community with a new tool for levitated  systems. The scalability of the underlining Maxwell equations means this phenomenon is not restricted to the nanoscale and so could potentially have an even wider impact than those envisaged in this work.


This work was supported by European Research Council Starting Grant ERC-2016-STG-714151-PSINFONI. We also thank A. Miroshnichenko for providing the material properties for the core-shell particle.


%

\end{document}


\begin{titlepage}
\thispagestyle{empty}

\title{Supplementary Information}
\author{Jack J. \surname{Kingsley-Smith}}
\author{Michela F. \surname{Picardi}}
\author{Francisco J. \surname{Rodr\'iguez-Fortu\~no}}
\affiliation{Department of Physics, King's College London, Strand, London WC2R 2LS, United Kingdom}
\date{\today}

\maketitle

\tableofcontents

\vfill

\end{titlepage}

\setcounter{equation}{0}
\renewcommand{\theequation}{S\arabic{equation}}
\setcounter{figure}{0}
\renewcommand{\thefigure}{S\arabic{figure}}

\section{\label{sec:exactforces}Dipole backscattering force theory}
\subsection{\label{subsec:exactmforce}Derivation of general expression}

Here, we derive the exact expression for the force on a near-surface electromagnetic dipole, with dipole moments $\mathbf{p}$ and $\mathbf{m}$, due to its own reflected scattering. We begin with the angular representation of the dipole's reflected fields \cite{Picardi2017} that are referred to in the main text, which we now explicitly show:
\begin{align}\label{eq:angularrep}
\mathbf{E}^{\text{ref}}(\mathbf{r}) = \iint \mathbf{E}^{\text{ref}}(k_x,k_y,z) \, e^{i\, (k_x\,x + \, k_y \, y)}\, dk_x \, dk_y, \nonumber \\
\mathbf{H}^{\text{ref}}(\mathbf{r}) = \iint \mathbf{H}^{\text{ref}}(k_x,k_y,z) \, e^{i\, (k_x\,x + \, k_y \, y)}\, dk_x \, dk_y.
\end{align}
The force can be expressed generally, for any electromagnetic dipolar source, from the following dipole force equation \cite{Chaumet2009,Kingsley-Smith2018}
\begin{equation}\label{eq:gradforce}
\langle \mathbf{F} \rangle = \frac{1}{2} \mathbb{R} \bigg\{ \big[\mathbf{\nabla} \otimes \mathbf{E}^{\text{ref}}(\mathbf{r})\big] \mathbf{p}^* + \mu_0 \big[\mathbf{\nabla} \otimes \mathbf{H}^{\text{ref}}(\mathbf{r})\big] \mathbf{m}^* - \frac{k_0^4}{6 \pi \varepsilon_0 c} (\mathbf{p} \times \mathbf{m}^*) \bigg\},
\end{equation}
where $\otimes$ represents an outer product, $c$ is the speed of light, $k_0$ is the wavenumber and $\varepsilon_0$ and $\mu_0$ are the permittivity and permeability of free space, respectively. 
By selecting the coordinate system such that the dipole location is $\mathbf{r} = (0,0,h)$, we can specify the reflected electromagnetic fields at the location of the dipole moments in terms of spatial frequencies \cite{Picardi2017,Kingsley-Smith2018}
\begin{align}\label{eq:AnguReprH}
\mathbf{E}_\mathbf{p}^{\text{ref}}(k_x,k_y,h) &=  \frac{i \, k_0^2}{8 \pi^2 \, \varepsilon_0 \, k_z} \big[r_p(\hat{\mathbf{e}}_p^- \cdot \mathbf{p})\hat{\mathbf{e}}_p^+ + r_s(\hat{\mathbf{e}}_s \cdot \mathbf{p})\hat{\mathbf{e}}_s\big] e^{2\,i\,h\,k_z}, \nonumber \\
\mathbf{E}_\mathbf{m}^{\text{ref}}(k_x,k_y,h) &=  \frac{i \, k_0^2}{8 \pi^2  \,\varepsilon_0 \, k_z c} \big[r_p(\hat{\mathbf{e}}_s \cdot \mathbf{m})\hat{\mathbf{e}}_p^+ - r_s(\hat{\mathbf{e}}_p^- \cdot \mathbf{m})\hat{\mathbf{e}}_s \big] e^{2\,i\,h\,k_z}, \nonumber \\
\mu_0\, \mathbf{H}_\mathbf{p}^{\text{ref}}(k_x,k_y,h) &=  \frac{i \, k_0^2}{8 \pi^2  \,\varepsilon_0 \, k_z c} \big[r_p(\hat{\mathbf{e}}_p^- \cdot \mathbf{p})\hat{\mathbf{e}}_s - r_s(\hat{\mathbf{e}}_s \cdot \mathbf{p})\hat{\mathbf{e}}_p^+ \big] e^{2\,i\, h\, k_z}, \nonumber \\
\mu_0\, \mathbf{H}_\mathbf{m}^{\text{ref}}(k_x,k_y,h) &=  \frac{i \, k_0^2}{8 \pi^2  \,\varepsilon_0 \, k_z c^2} \big[r_p(\hat{\mathbf{e}}_s \cdot \mathbf{m})\hat{\mathbf{e}}_s + r_s(\hat{\mathbf{e}}_p^- \cdot \mathbf{m})\hat{\mathbf{e}}_p^+ \big] e^{2\,i\,h\,k_z}.
\end{align}
where $\mathbf{E}^{\text{ref}} = \mathbf{E}_\mathbf{p}^{\text{ref}} + \mathbf{E}_\mathbf{m}^{\text{ref}}$ and $\mathbf{H}^{\text{ref}} = \mathbf{H}_\mathbf{p}^{\text{ref}} + \mathbf{H}_\mathbf{m}^{\text{ref}}$. 
We are using the $s$ and $p$ polarization basis vectors defined in Refs \cite{Rotenberg2012, Picardi2017} as $\hat{\mathbf{e}}_s = (k_x^2+k_y^2)^{-\frac{1}{2}}(-k_y \hat{\mathbf{x}} + k_x \hat{\mathbf{y}})$ and $\hat{\mathbf{e}}_p^\pm = \hat{\mathbf{e}}_s \times \frac{\mathbf{k}^\pm}{k_0}$, respectively. This basis is convenient because we can immediately see that, for example, $\mathbf{m} = (0,0,m_z)$ leads to the first term of $\mathbf{H}_\mathbf{m}^{\text{ref}}$ vanishing and with it, any $p$-polarized reflection. Likewise, $\mathbf{p} = (0,0,p_z)$ makes the second term in $\mathbf{E}_\mathbf{p}^{\text{ref}}$ and removes the $s$-polarized reflection.

The full expression for the backscattering force on a source with any $\mathbf{p}$ and $\mathbf{m}$ is given by substituting (\ref{eq:angularrep}) and (\ref{eq:AnguReprH}) into (\ref{eq:gradforce}) to form
\begin{equation}\label{eq:mxmymzforce}
    \langle F_z \rangle =  \frac{1}{16 \pi}\mathfrak{Re}  \bigg\{\int_0^\infty e^{2\, i \, h \, k_{z}} (\mathcal{F}_z^{p_{xy}} + \mathcal{F}_z^{p_{z}} + \mathcal{F}_z^{m_{xy}} + \mathcal{F}_z^{m_{z}}) \, dk_t \bigg\}  ,
\end{equation}
\begin{align*}
    \mathcal{F}_z^{p_{xy}} &\equiv -\frac{2}{\varepsilon_0}  \, \big(|p_x|^2 + |p_y|^2\big) \, k_t \,  \Big(k_{z}^2 \, r_p - k_0^2 \, r_s\Big),\\
    \mathcal{F}_z^{p_{z}} &\equiv -\frac{2}{\varepsilon_0} \, |p_z|^2 \, k_t^3 \,  r_p,\\
    \mathcal{F}_z^{m_{xy}} &\equiv \mu_0 \, \big(|m_x|^2 + |m_y|^2\big) \, k_t \, \Big(k_{z}^2 \, r_s - k_0^2 \, r_p\Big),\\
    \mathcal{F}_z^{m_{z}} &\equiv -2 \, \mu_0\, |m_z|^2 \, k_t^3 \,  r_s.
\end{align*}

\subsection{\label{subsec:PEC}Derivation of force over a PEC and applying image theory}

By applying the PEC condition ($r_s = -1$) to the vertical magnetic dipole (VMD) case in Eq.\,(\ref{eq:AnguReprH}), one can clearly see how the reflected fields become identical to that of an image VMD $\mathbf{m}' = (0,0,m_z')$ located at a distance $2h$ below the source VMD,
\begin{equation}\label{eq:imagetheory}
\mathbf{H}^{\text{ref}}(k_x,k_y,h) =  \frac{i \, k_0^2}{8 \pi^2 \, k_z} \Big[\big(\hat{\mathbf{e}}_p^- \cdot \mathbf{m}'\big)\hat{\mathbf{e}}_p^+ \Big] e^{2\, i\, h\,k_z}.
\end{equation}
The equivalence can be simply described by $m_z' = r_s \, m_z$, where $r_s = -1$. The consequent force between the source and image VMDs is given by substituting (\ref{eq:imagetheory}) into (\ref{eq:gradforce}),
\begin{equation}\label{eq:PEC}
    \evalat{\langle F_z \rangle^{\text{PEC}}}{\theta = 0^\circ} = m_z^* \, m_z'\,  \frac{\mu_0}{64 \, \pi\, h^4}\, \mathfrak{Re}\Bigg\{\Big(4 \, h^2  \, k_0^2 + 6\,i\, h\, k_0 -3\Big) \, e^{2\,i\,h\,k_0} \Bigg\}.
\end{equation}
This expression is valid for all values of $h$ and the difference in signs between $m_z$ and $m_z'$ ensures that the $\langle F_z \rangle>0$ (repulsive) for small values of $h$. In the limiting case where $h\to0$, 
\begin{equation}\label{eq:VMDQSPECforce}
    \lim_{h\to0} \evalat{\langle F_z \rangle^{\text{PEC}}}{\theta = 0^\circ} =  -m_z^* \, m_z' \, \frac{3 \, \mu_0}{64 \,\pi \, h^4}.
\end{equation}

We now consider the horizontal magnetic dipole (HMD) case where $\theta = 90^\circ$ which, unlike the VMD, involves both $p$ and $s$-polarized backscattering. Following the same procedure as before and applying $r_p = 1$, we arrive at
\begin{equation}
    \evalat{\langle F_z \rangle^{\text{PEC}}}{\theta = 90^\circ} = m_t^* \, m_t'\,\frac{\mu_0}{128\,\pi\, h^4} \mathfrak{Re}\Bigg\{\Big(3 - 6\,i\,h\,k_0 - 8\, i\,h^2\, k_0^2 + 8\,i\,h^3\,k_0^3\Big) \, e^{2\,i\,h\,k_0} \Bigg\},
\end{equation}
where the transverse dipole moment $m_t = \sqrt{m_x^2 + m_y^2}$, which is applicable because of the rotational symmetry of the problem. The image horizontal dipole moment $m_t'$ is equal to the HMD moment in the PEC limit \cite{Balanis1997} (i.e. $m_t' = m_t$). By taking the same $h\to0$ limit as before, we arrive at a quasistatic PEC force equation which behaves the same as Eq.\,(\ref{eq:VMDQSPECforce})
\begin{equation}
    \lim_{h\to0} \evalat{\langle F_z \rangle^{\text{PEC}}}{\theta = 90^\circ} =  m_t^* \, m_t' \, \frac{3 \, \mu_0}{64 \,\pi \, h^4},
\end{equation}
indicating that the orientation of the magnetic dipole does not affect its near-field repulsion from a PEC substrate. The influence of this invariance is apparent in the realistic metal substrate too, which we discuss in Section \ref{subsec:Rotation}.

\subsection{\label{subsec:UltraNF}Transitioning from a PEC to a realistic metal}

In the main text, we described how the repulsion of a VMD undergoes a $h$ dependence transition depending on how closely the substrate's optical response resembles that of a PEC. For realistic metals with a finite permittivity and skin depth, the reflection coefficient $r_s$ tends to $0$ for high values of $k_t$ and one must expand $r_s$ around $k_t\gg k_0$ to find non-zero higher order terms
\begin{equation}\label{eq:rsexpansion}
    r_s = 0 + k_0^2 \, \frac{\varepsilon_{\text{s}} - 1}{4 \, k_{t}^2}  + \mathcal{O}\bigg(\frac{1}{k_{t}^6}\bigg).
\end{equation}
\noindent When dominant, these higher orders terms alter the form of the integrand $\mathcal{F}_z^{m_z}$ in Eq.\,(\ref{eq:mxmymzforce}) and consequently changing the $h$ dependence. For the VMD, the force integral, which includes $k_t^3 \, r_s$, takes the form of 
\begin{equation}\label{eq:integralform}
    \int_0^\infty e^{i\,a \sqrt{1-x^2}}\, x^b \, dx
\end{equation}
where $b$ is an integer that changes depending on which term in (\ref{eq:rsexpansion}) is dominant and $a<4\pi$ when $h<\lambda$. Unfortunately, (\ref{eq:integralform}) is only analytically solvable when $b>0$ and odd. The $b = 3$ case corresponds to the PEC case when $r_s = -1$ and gives Eq.\,(\ref{eq:PEC}) its $h^{-4}$ dependence. The $b=1$ case corresponds to the first non-zero term of (\ref{eq:rsexpansion}) and leads to the $h^{-2}$ dependent quasistatic force Eq.\,(2) in the main text. Higher order terms in (\ref{eq:rsexpansion}) would correspond to cases where $b<0$ and so are not explicitly solvable. 

\begin{figure}
    \centering
    \includegraphics[width=0.6\textwidth]{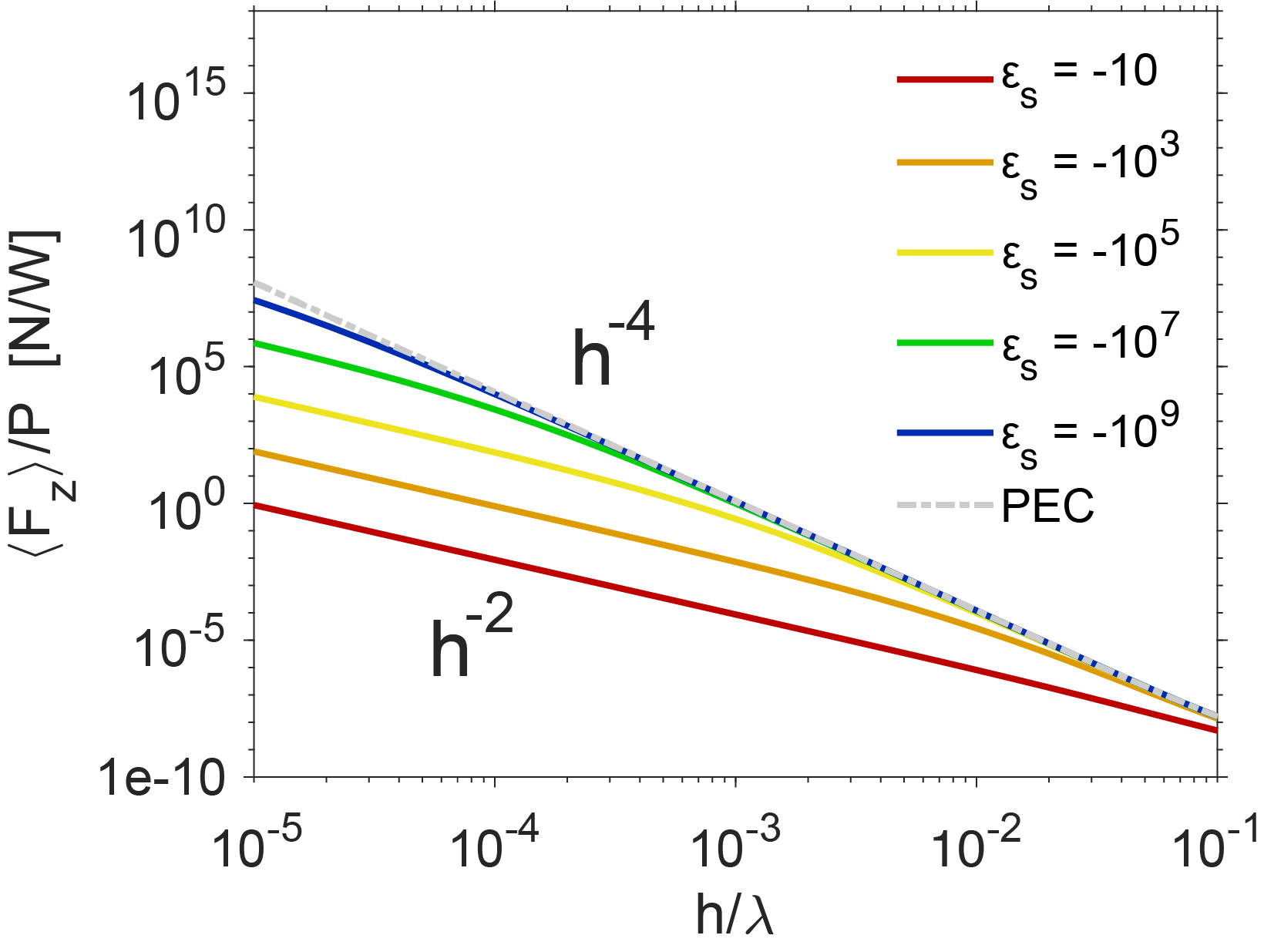}
    \caption{The height dependence of the normalized vertical force on a point VMD near substrates of varying permittivities. For large negative $\varepsilon_s$, the dipole is repelled as if it was a PEC substrate, until it reaches a height comparable to the skin depth, at which point the PEC model ($r_s = -1$) breaks down, and the quasistatic model becomes dominant and the height dependence switches from $h^{-4}$ to $h^{-2}$.}
    \label{fig:ultraclose}
\end{figure}
When conducting these calculations for various surfaces and extending the domain of $h$ down to extremely small distances, as is shown in Fig. \ref{fig:ultraclose}, we see that only surfaces with a large negative $\varepsilon_s$ exhibit the $h^{-4}$ dependence of the PEC case and this gradually transitions towards the $h^{-2}$ dependence of Eq.\,(3) of the main text. 
Prior works that employ image theory to explain levitation \cite{Rodriguez-Fortuno2014} show a constant $h$ dependence in the quasistatic regime. It is therefore striking that we observe a transition in this case and it is an unexpected consequence of the higher order terms of $r_s$ altering the nature of the interaction.

\subsection{\label{subsec:Rotation}Force on a rotated magnetic dipole}

The theoretical analysis in the main text was restricted to the case of the exact VMD. This configuration was convenient because the VMD lacks any $p$-polarized scattering and the repulsive force is due to the $s$-polarized reflection. Since an experimental realization, with its corresponding limitations, may not achieve a \textit{perfectly} vertical orientation, it is helpful to gauge the robustness of this $\mathfrak{Re}\{\varepsilon_{\text{s}}\}<1$ repulsion under dipole rotations.

\begin{figure*}[t]
    \centering
    \includegraphics[width=\textwidth]{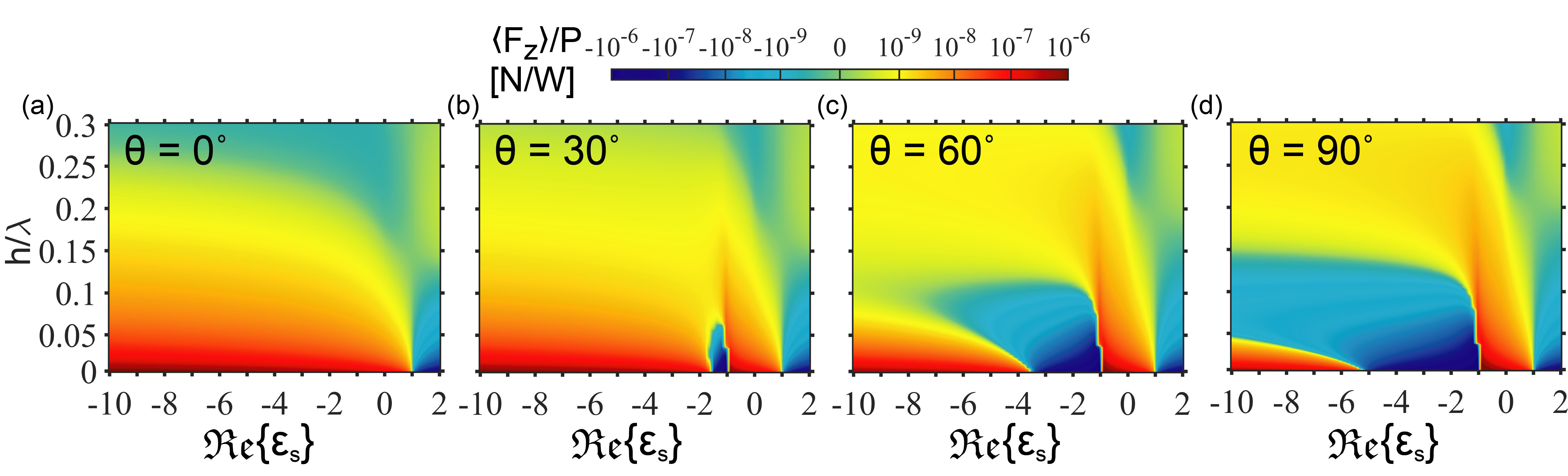}
    \caption{(a)-(d) As the VMD is rotated by the angle $\theta$, the time-averaged $\mathfrak{Re}\{\varepsilon_{\text{s}}\}<1$ repulsion is partially disrupted by a new attractive region. The attraction is driven by the $p$-polarized backscattering of the dipole and in the extreme case of the horizontal dipole, the repulsion remains dominant when $\mathfrak{Re}\{\varepsilon_{\text{s}}\} \lessapprox -5$ and $h$ is small enough. Note that the repulsion behavior is only disrupted for relatively large values of $\theta$ and so an approximately vertical dipole will experience the same force as the ideal case.}
    \label{fig:rotation}
\end{figure*}

Fig. \ref{fig:rotation}a shows that rotating the dipole by small angles has no appreciable effect on the nature of the force. As the angle increases towards a HMD, a new attractive region appears near $\mathfrak{Re}\{\varepsilon_{\text{s}}\} = -1$ owing to the $p$-polarized surface plasmon resonance of the material. This signifies a competition between the attractive $p$ and repulsive $s$-polarized responses that depends strongly on $\mathfrak{Re}\{\varepsilon_{\text{s}}\}$. The HMD has the strongest $p$-polarization scattering of any magnetic dipole orientation and so for an experimental realization, this represents the worst case scenario for observing this effect. However Fig. \ref{fig:rotation}d clearly shows that, even in that worst case, given $\mathfrak{Re}\{\varepsilon_{\text{s}}\} \lessapprox -5$, there exists a $h$ near the surface that will experience a repulsion. Given that most common metals fulfill this requirement, we therefore claim that the magnetic dipole repulsion is robust to significant rotation, assuming an appropriate surface material. The repulsion of the HMD is explained by means of image theory in Section \ref{subsec:PEC}.

We note that extending this study to elliptical dipoles does not yield any new physics because the vertical force is invariant to the phases of the dipole moments, as can be seen from Eq.\,(\ref{eq:mxmymzforce}). Diagonal and elliptical dipoles experience the same vertical forces and are merely linear combinations of the VMD and HMD. 

\section{\label{sec:MST}Maxwell stress tensor calculations}

The Maxwell stress tensor $\overset\leftrightarrow{\mathbf{T}}$ is a widely used technique in optics for determining the optical force on any body. The second rank tensor is derived from the flow of electromagnetic momentum through an arbitrary surface and is related to the time-averaged optical force by the following surface integral \cite{Novotny2006}
\begin{equation}\label{equ:MSTsurf}
\langle\mathbf{F}\rangle = \int_{\mathcal{S}}^{} \langle\overset\leftrightarrow{\mathbf{T}}\rangle \cdot \textbf{\^n} \, d\mathcal{S}
\end{equation}
\noindent where $\mathbf{F}$ is the force acting on a body and $\textbf{\^n}$ is the normal vector perpendicular to and out of any arbitrary closed surface $\mathcal{S}$ enclosing the body. The angular bracket notation around a single variable indicates a time averaging. $\langle\overset\leftrightarrow{\mathbf{T}}\rangle$ is defined as \cite{Novotny2006}
\begin{equation}
\langle\overset\leftrightarrow{\mathbf{T}}\rangle = \frac{1}{2} \mathbb{R} \bigg\{\varepsilon \mathbf{E} \otimes \mathbf{E}^* + \mu \mathbf{H} \otimes \mathbf{H}^* - \frac{1}{2} \big(\varepsilon |\mathbf{E}|^2 + \mu |\mathbf{H}|^2\big)\overset\leftrightarrow{\mathbf{I}}\bigg\}
\label{equ:MST}
\end{equation}
\noindent where $\mathbf{E}$ and $\mathbf{H}$ are the total electric and magnetic fields, $\otimes$ denotes the outer product of two vectors, asterisks represent complex conjugations, $\overset\leftrightarrow{\mathbf{I}}$ is the identity matrix and $\varepsilon$ and $\mu$ are the permittivity and permeability of the medium, respectively. 

In Fig. 1d of the main text, the electromagnetic fields were calculated analytically from Green's theorem for a dipole over a surface \cite{Novotny2006}, shown in Eq. \ref{eq:AnguReprH}. For the core-shell system considered later in the main text, $\mathbf{E}$ and $\mathbf{H}$ were calculated in the time-domain with the commercial simulation software \textit{CST Microwave Studio}. The force calculations of Fig. 2b  were conducted in much the same way as Ref. \cite{Rodriguez-Fortuno2015}. That is, the particle is illuminated with a plane wave and the total fields are calculated. The simulation is then repeated with the same mesh profile but without the particle. The fields from the latter simulation are then subtracted from the total fields to produce the scattered fields of the particle only, without the illumination. This allows us to isolate the the particle's backscattering force from the radiation pressure of the illumination and compare directly with the general dipole theory, $F_z^{\text{th}}$. 

We provide below in Fig.\,\ref{fig:MSTtotalfield} the force calculated from the total fields, which includes the incident plane wave. This total force $F_z^{\text{tot}}$ corresponds to that what would be observed in an experiment. We note that the repulsion is still strongest around the magnetic resonance at  1.3 $\mu$m, and that this data highlights how strong the backscattering force is. The isolation of the backscattering force was presented in Fig.\,2 of the main text because it allowed for a direct comparison with the general backscattering dipole force theory. The stress tensor calculations included varying the integration cube sizes to check for convergence. 
\begin{figure}[h]
    \centering
    \includegraphics[width=0.65\textwidth]{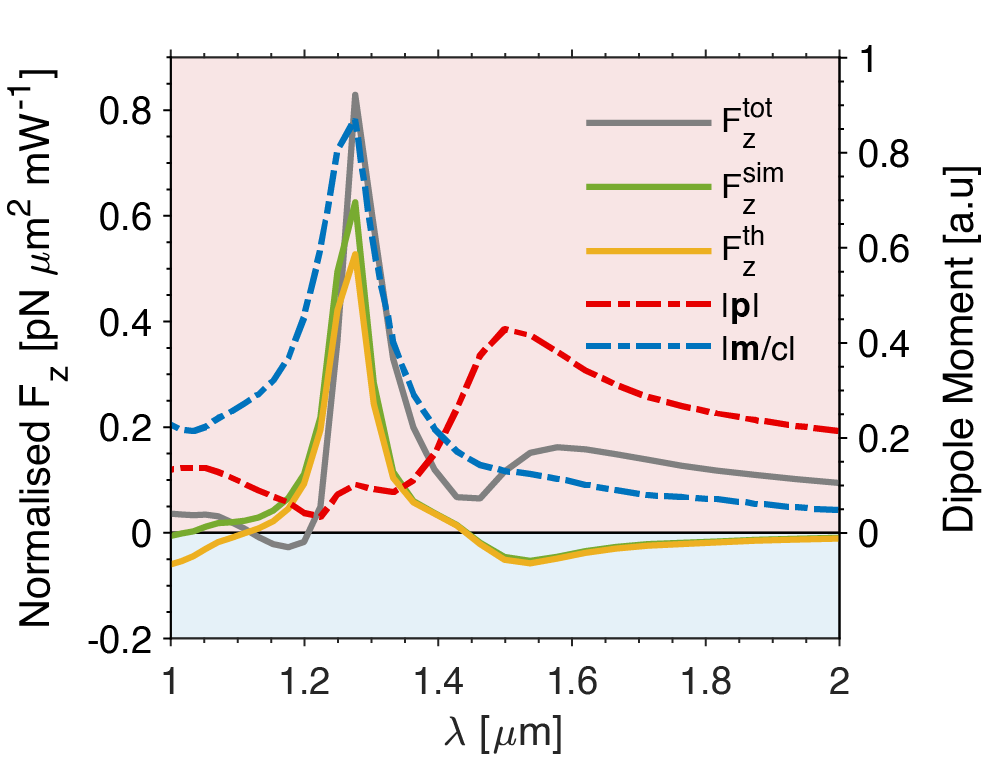}
    \caption{Wavelength dependence of $F_z^{\text{tot}}$, calculated from the total fields (including the plane wave), plotted alongside the retrieved dipole moments and backscattering force results from Fig.\,2 in the main text.}
    \label{fig:MSTtotalfield}
\end{figure}

\section{\label{sec:DipoleRetrieval}Dipole retrieval from a source near a surface}

There are many methods for determining the induced electric and magnetic dipole resonances of an illuminated structure. Mie theory \cite{Bohren1998} is an exact solution for this problem if spherical symmetry is conserved but the presence of the surface in our current problem breaks this symmetry. Some methods also lack the ability to determine the orientation of the induced dipoles or the relative phase between their components. Rather than working with a weak assumption that the surface effects were negligible, we opted for our own method based on the decomposition of the scattered far fields into orthogonal basis functions. This method is valid for any system which can be described exactly with the angular representation and can be easily extended to higher order multipoles \cite{Vazquez-Lozano2019}. 

\subsection{\label{subsec:decompositionalgebra}Decomposition of radiation pattern into orthogonal basis functions}

We look to expand the numerically calculated radiation pattern $E_{\text{pattern}}$ in such a way that the complex Cartesian dipole moments are weighting coefficients
\begin{equation}\label{eq:DipoleBasis}
    E_{\text{pattern}} = \, p_x \, \hat{E}_{p_x} + p_y \, \hat{E}_{p_y} +p_z \, \hat{E}_{p_z} + \frac{m_x}{c} \, \hat{E}_{m_x} + \frac{m_y}{c} \, \hat{E}_{m_y}+ \frac{m_z}{c} \, \hat{E}_{m_z}.
\end{equation}
\noindent where $\hat{E}_{i}$ is a basis function corresponding to the radiation pattern of the $i^{\text{th}}$ dipole moment. For the purpose of our mathematical calculation, we take $p_i$ and $m_i$ to be dimensionless. However, these parameters gain units by comparing them with the dipole moments used to create the basis functions $\hat{E}$, which have units. 

The retrieval method is reliant on a single condition; the basis functions must be orthogonal to each other. The far-field radiation diagrams for different multipoles would be exactly orthogonal in the case when the source was in free space, because they correspond to orthogonal vector spherical harmonics. However, we found out that when accounting for the surface reflection, the far-field radiation of the multipoles above the surface are no longer orthogonal to each other. Therefore, we had to apply an orthogonalization procedure enforced in Section \ref{subsec:definedipoleangularspec}. 
Using this orthogonal basis, 
\begin{equation}\label{eq:orthogonalbasissum}
E_{\text{pattern}} = \sum_{i=1}^6 a_i \, \hat{E}_i
\end{equation}
where the coefficients $a_i = \langle \hat{E},\hat{E}_i\rangle / \langle \hat{E}_i, \hat{E}_i \rangle$ and the angular bracket notation around two variables corresponds to their overlap integral, defined as $\langle E_1,E_2 \rangle = \iint \mathbf{E}_1 \cdot \mathbf{E}_2^* \, d\Omega$, where $\Omega$ is the solid angle and the integration is conducted over the top hemisphere (i.e. far field radiation over the surface). A transformation is then performed to move from the orthogonalized basis to the basis of dipole moment of Eq.\,(\ref{eq:DipoleBasis}).

\subsection{\label{subsec:deriveEpattspectrum}Relating the angular spectrum representation to the far-field pattern}

To obtain the radiation diagrams of the unit dipoles above a surface, we will rely on the angular spectrum representation of the scattered fields of a dipole near a surface, which are known analytically via Green's function method for a dipole above a surface. These angular spectra can then be converted into the far-field distance-independent pattern $\mathbf{E}_\text{pattern}(\theta,\phi)$.
That is to say, the exact fields of a dipole in Fourier space can be converted into a distance-independent radiation pattern which commercial software can readily compute for any physical scenario.
$\mathbf{E}_\text{pattern}$ is related to the far fields $\mathbf{E}_\text{far}$ by
\begin{equation}\label{eq:epattern}
    \mathbf{E}_\text{far} (\mathbf{r}) = \mathbf{E}_\text{pattern} (\theta,\phi) \, \frac{e^{i \, k_0 \, r}}{r}.
\end{equation}
We write the real-space fields here in spherical coordinates to align with the convention in radiation patterns, where $\mathbf{r} = (r,\theta,\phi)$ is the general position vector in spherical coordinates. The angular representation derives from the Weyl identity
\begin{equation}\label{eq:weyl}
    \frac{e^{i \, k_0 \, r}}{r} = \iint \frac{i \, k_0}{2 \pi} e^{i \, \mathbf{k} \cdot \mathbf{r}} d\Omega.
\end{equation}
The definition of the angular representation itself is
\begin{equation}\label{eq:angularspectrum}
    \mathbf{E}(\mathbf{r}) = \iint k_0 \, k_z \, \mathbf{E}(k_x,k_y;0) e^{i \, \mathbf{k} \cdot \mathbf{r}} d\Omega. 
\end{equation}
Note that this definition differs to that of Eq.\,(\ref{eq:angularrep}) where the integrals are in terms of $dk_x$ and $dk_y$. Here, we have transformed the integrals in (\ref{eq:weyl}) and (\ref{eq:angularspectrum}) into the solid angle $\Omega$ of a sphere. 

Our aim is to find a clear relation between $\mathbf{E}_\text{pattern}(\theta,\phi)$ and $\mathbf{E}(k_x,k_y;0)$. To do this, we can find the relation for a simple example. In the far field, the far field pattern of a VED is $\mathbf{E}_\text{far} \propto \text{sin}(\theta) \frac{e^{i \, k_0 \, r}}{r}$ and therefore $\mathbf{E}_\text{pattern} \propto \text{sin}(\theta)$. Substituting these relations into Eq.\,(\ref{eq:weyl}) leads to 
\begin{equation}\label{eq:subVED}
    \text{sin}(\theta) \frac{e^{i \, k_0 \, r}}{r} = \iint \frac{k_t}{k_0} \frac{i \, k_0}{2 \pi} e^{i \, \mathbf{k} \cdot \mathbf{r}} d\Omega = \iint \frac{i \, k_t}{2 \pi} e^{i \, \mathbf{k} \cdot \mathbf{r}} d\Omega.
\end{equation}
Comparing (\ref{eq:subVED}) with (\ref{eq:angularspectrum}) leads us to the desired transformation in a generalized form which will be true in general for any point source
\begin{equation}\label{eq:Epatterntoangularspectrum}
    \mathbf{E}_\text{pattern}(\theta,\phi) = -2\, \pi \,i \, k_z \, \mathbf{E}(k_x,k_y;0).
\end{equation}

\subsection{\label{subsec:definedipoleangularspec}Orthogonal basis functions in angular representation}

The angular representation of the exact scattered fields of an electric or magnetic dipole, at $z = h$, with a reflecting surface at $z=0$ are known \cite{Picardi2017},
\begin{align}\label{eq:angularspectraspherical}
     \mathbf{E}(k_x,k_y;0) &=  \begin{pmatrix}
            E_k\\
            E_s\\
            E_p
        \end{pmatrix} \xLongrightarrow{\text{far-field}} \begin{pmatrix}
            E_r\\
            E_\phi\\
            E_\theta
        \end{pmatrix} = \begin{pmatrix}
            0\\
            E^\text{self}_{\phi} + E^\text{ref}_{\phi}\\
            E^\text{self}_{\theta} + E^\text{ref}_{\theta}
        \end{pmatrix}
    \nonumber \\
     E^\text{self}_{\phi}(k_x,&k_y;0) = \frac{i \, k_0^2}{8 \pi^2 \varepsilon_0 k_z} \, \Big(\hat{\mathbf{e}}_s \cdot \mathbf{p} -\hat{\mathbf{e}}_p^+ \cdot \frac{\mathbf{m}}{c}\Big) \, e^{-i k_z h} \nonumber \\
     E^\text{ref}_{\phi}(k_x,&k_y;0) = \frac{i \, k_0^2}{8 \pi^2 \varepsilon_0 k_z} \,r_s \, \Big(\hat{\mathbf{e}}_s \cdot \mathbf{p} - \hat{\mathbf{e}}_p^- \cdot \frac{\mathbf{m}}{c}\Big) \, e^{i k_z h}. \nonumber \\
     E^\text{self}_{\theta}(k_x,&k_y;0) = \frac{i \, k_0^2}{8 \pi^2 \varepsilon_0 k_z} \, \Big(\hat{\mathbf{e}}_p^+ \cdot \mathbf{p} + \hat{\mathbf{e}}_s \cdot \frac{\mathbf{m}}{c}\Big) \, e^{-i k_z h} \nonumber \\
     E^\text{ref}_{\theta}(k_x,&k_y;0) = \frac{i \, k_0^2}{8 \pi^2 \varepsilon_0 k_z} \,r_p \, \Big(\hat{\mathbf{e}}_p^- \cdot \mathbf{p} + \hat{\mathbf{e}}_s \cdot \frac{\mathbf{m}}{c}\Big) \, e^{i k_z h} \nonumber \\
\end{align}
\noindent When only considering the electric and magnetic dipole, one can form six basis functions by simply setting all but one component of the electromagnetic dipole to zero and substituting  (\ref{eq:angularspectraspherical}) into (\ref{eq:Epatterntoangularspectrum}). Thus the first basis function could be formed from $(p_x,p_y,p_z,m_x,m_y,m_z) \rightarrow{} (1,0,0,0,0,0)$. 

Though we have now formed six basis functions, they do not yet necessarily form an orthogonal set. To meet the orthogonalization condition outlined in Subsection \ref{subsec:decompositionalgebra}, we can apply the Gram-Schmidt procedure
\begin{align}
    \mathbf{u}_1 &= \mathbf{v}_1 \nonumber \\
    \mathbf{u}_k = \mathbf{v}_k - &\sum_{j=1}^{k-1}\frac{\langle\mathbf{v}_k,\mathbf{u}_j\rangle}{\langle\mathbf{u}_j,\mathbf{u}_j\rangle}\mathbf{u}_j,
\end{align}
\noindent where $\mathbf{u}$ is a general basis vector from a non-orthogonal set which is orthogonalized to form $\mathbf{v}$ which forms an orthogonal set of vectors. This enables us to obtain an orthogonal basis (as in \ref{eq:orthogonalbasissum}) from the original dipolar basis (\ref{eq:DipoleBasis}). The new set of orthogonal radiation pattern basis functions become combinations of dipole moments which are linearly independent from one another.

\bibliography{SI.bbl}